\documentstyle[11pt]{article}

\skip\footins 5mm plus 2mm minus 2mm
\begin{document}
\setlength{\parindent}{0 mm}

\vspace*{-0.5in}

\vspace*{-0.5in}

\def\bd{\begin{displaymath}}
\def\ed{\end{displaymath}}
\def\be{\begin{equation}}
\def\ee{\end{equation}}
\def\p{\partial}

\newtheorem{theorem}{Theorem}
\newtheorem{conjecture}{Conjecture}
\newtheorem{lemma}{Lemma}
 \newtheorem{proposition}{Proposition}
\newcommand {\ds} {\displaystyle \sum}
\vspace*{-0.5in}

\renewcommand{\thesection}{\arabic{section}}

\large
\vskip 1 cm
\vskip 1cm 
\vskip 1cm

\centerline { {\bf  BOGOYAVLENSKY--VOLTERRA AND  }  }
\centerline { {\bf     BIRKHOFF INTEGRABLE SYSTEMS}  }

 \vskip 1  cm
\centerline { Pantelis A. Damianou and Stelios P. Kouzaris}

\vskip .5 cm

 \centerline { Department of
Mathematics and Statistics } \centerline  { University of Cyprus}

\centerline { P. O. Box 20537, 1678 Nicosia, Cyprus } 
\centerline{Email: damianou@ucy.ac.cy, \ \ \ maskouz@ucy.ac.cy}
 \vskip 2 cm \centerline { \bf {ABSTRACT }}
\bigskip
{\it  In this paper we examine an interesting connection between the generalized Volterra lattices of Bogoyavlensky
and a special case of an integrable system defined by Sklyanin. The Sklyanin system happens to be
one of the cases  in the classification of Kozlov and Treshchev of Birkhoff integrable Hamiltonian systems. Using this
connection we  demonstrate  the integrability of the system and define a new Lax pair representation.
 In addition, we comment on  the bi--Hamiltonian  structure of the system.  }

\newpage

\section{Introduction--Birkhoff Integrable systems}

In this paper we examine an interesting connection between the generalized Volterra lattices of Bogoyavlensky \cite{bogo2}
and a special case of an integrable system defined by Sklyanin \cite{sklyanin}. The Sklyanin system happens to be
one of the systems in the classification of Kozlov and Treshchev of Birkhoff integrable Hamiltonian systems. Using this
connection we are able to prove the integrability of the system and define a new Lax pair representation different from
the one in \cite{ranada}. In addition, we comment on  the multi--Hamiltonian  structure of the system.
This connection was discovered in an effort to connect the Volterra $D_n$ system with the corresponding Toda $D_n$ system
as in the case of the other classical Lie groups.  In contrast to the other cases,  the
Volterra $D_n$ system does not correspond to the Toda $D_n$ system under the procedure of Moser but it actually
 corresponds to a special case of the  Sklyanin system mentioned above. 
 A Miura type transformation between the Volterra and Toda $D_n$ systems is therefore still an open problem.

We begin with the following more general definition which involves systems with exponential 
interaction: Consider a Hamiltonian of the form
\be
H={1 \over 2} ({\bf p}, {\bf p})+\sum_{i=1}^N e^{ ({\bf v}_i\,,\, {\bf q}) }  \ , \label{a1} 
\ee
where ${\bf q}=(q_1, \dots, q_n)$,  ${\bf p}=(p_1, \dots, p_n)$, ${\bf v}_1, \dots, {\bf v}_N$ are vectors in ${\bf R}^n$ and $(\ , \ )$ is the standard
inner product in ${\bf R}^n$.
 The set of vectors  $\Delta=\{ {\bf v}_1, \dots, {\bf v}_N \}$ is  called the spectrum of the system.

Let $M$ be the $N \times N$ matrix whose elements are 
\bd
M_{ij}=\left( {\bf v}_i, {\bf v}_j \right) \ .
\ed
Hamilton's equations  of motion can be transformed by a generalized Flaschka transformation  to a polynomial system of $2N$ differential equations. 
The transformation is defined as follows:

\be
a_i=-e^{ ({\bf v}_i  , {\bf q})} ,  \qquad b_i=( {\bf v}_i, {\bf p}) \label{a66}  \ .
\ee
 We end--up with a system of polynomial differential  equations
\begin{eqnarray}
\dot{a}_k&=&a_k b_k \nonumber  \\
\dot{b}_k&=&\sum_{i=1}^N M_{ki} a_i  \label{a99} \ .
\end{eqnarray}
Equations (\ref{a99}) admit the following two integrals
\be
F_1=\sum_{i=1}^N \lambda_i b_i  \ , \qquad  F_2=\prod_{i=1}^N a_i^{\lambda_i}  \label{a98} 
\ee
provided that  there exist constants $\lambda_i$ such that $\sum_{i=1}^N \lambda_i {\bf v}_i=0$. Such integrals always exist for $N>n$.
One can define a  canonical bracket  on the space of variables $(a_i, b_i)$ by the formula
\bd
\{b_i, a_j \}=({\bf v}_i, {\bf v}_j ) a_j 
\ed
and all other brackets equal to zero. The integrals $F_1$ and $F_2$ are Casimirs of system (\ref{a99}).

An interesting case of (\ref{a1}) occurs  when  the spectrum is a system of simple roots for a simple Lie algebra
 ${\cal G}$. In this case $N=l={\rm rank} \,  {\cal G}$. It is worth mentioning that the case where $N, n$ are arbitrary is
an open and unexplored area of research. The main exception is the work of Kozlov and Treshchev \cite{kozlov}  where
a classification of system (\ref{a1}) is performed under the assumption that the system possesses $n$  polynomial (in the
momenta) integrals.  We also note the papers by Ranada  \cite{ranada},   Annamalai,  Tamizhmani  \cite{anna} and
Emelyanov \cite{emelyanov}.
 Such systems are called Birkhoff integrable. For each  Hamiltonian in (\ref{a1}) we 
associate a Dynkin type diagram as follows: It is a graph whose vertices correspond to the elements of  $\Delta$. Each pair of  vertices
${\bf v}_i$, ${\bf v}_j$ are connected by 
\bd
{ 4 ({\bf v}_i, {\bf v}_j)^2 \over ({\bf v}_i,{\bf v}_i) ({\bf v}_j, {\bf v}_j) }
\ed
edges.  

\bigskip
\noindent
{\bf Example:}
The classical  Toda lattice corresponds to a Lie algebra of type $A_{n-1}$. In other words $N=l=n-1$
 and we choose $\Delta$ to
be the set:
\bd
{\bf v}_1=(1,-1,0, \dots, 0),   \dots \dots \dots , {\bf v}_{n-1}=(0,0, \dots,0, 1,-1) \ .
\ed
The graph  is the usual Dynkin diagram of a Lie algebra of type $A_{n-1}$.
The Hamiltonian becomes:
\be
  H(q_1, \dots, q_n, \,  p_1, \dots, p_n) = \sum_{i=1}^n \,  { 1 \over 2} \, p_i^2 +
\sum _{i=1}^{n-1} \,  e^{ q_i-q_{i+1}}  \ , \label{a2}
\ee  
which is the well-known classical, non--periodic Toda lattice.  This system was investigated in 
  \cite{flaschka1},  \cite{flaschka2},  \cite{henon},  \cite{manakov}, \cite{moser},  \cite{toda}
and numerous  other papers that are impossible to list here.
This type of Hamiltonian was discovered   by Morikazu Toda  \cite{toda}. The original Toda lattice can be
viewed as a discrete version of the Korteweg--de Vries equation. It is called a lattice as in atomic lattice since
interatomic interaction was studied. This  system  appears also   in Cosmology,
in the work of Seiberg and Witten on supersymmetric Yang--Mills theories and it has applications
in analog computing and numerical computation of eigenvalues. But the Toda lattice is mainly a 
theoretical mathematical model which is important due to the rich mathematical structure encoded in it.
The Toda lattice is integrable in the sense of Liouville. There exist $n$ independent integrals of motion
in involution. These integrals are polynomial  in the momenta.

As we mentioned earlier, the  Toda lattice was generalized to the case where the spectrum corresponds to a root space of 
an arbitrary simple Lie group.
 These  systems generalize the usual finite, non--periodic
Toda lattice (which corresponds to a root system of type $A_n$). This 
generalization is due to Bogoyavlensky \cite{bogo3}. These systems were studied extensively in 
 \cite{kostant} where the solution of the systems  was connected intimately with the representation
theory of simple Lie groups. There are also studies by 
 Olshanetsky and Perelomov
\cite{olshanetsky} and Adler, van Moerbeke \cite{avm}.

It is  more convenient to work, instead with   the space of  the natural $(q,p)$ variables, with the 
Flaschka variables $(a,b)$ which are defined by:

\be
\begin{array}{lcl}
a_i& =&{1\over 2}e^{ {1\over 2} ({\bf v}_i, {\bf q})} \ \ \ \ \ i=1,2, \dots, N    \\
 b_i & = &-{ 1 \over 2} p_i  \ \ \ \ \ \ \ \ \ \ \ \ \  i=1,2, \dots, n \ .     \label{a3}
\end{array}
\ee

We end--up with a new set of polynomial  equations in the variables $(a,b)$.  One can write the equations
in Lax pair form, see for example \cite{perelomov}.
 The Lax pair ($L(t), B(t)$) in ${\cal G}$ can be described in terms
of the root system as follows:

\bd
L(t)=\sum_{i=1}^l b_i(t) h_{\alpha_i} + \sum_{i=1}^l a_i(t) (e_{\alpha_i}+e_{-\alpha_i}) \ ,
\ed

\bd
B(t)=\sum_{i=1}^l a_i(t) (e_{\alpha_i}-e_{-\alpha_i})  \ .
\ed
As usual $h_{\alpha_i}$ is an element of  a fixed Cartan subalgebra and $e_{\alpha_i}$ is a  root vector corresponding to
the simple root $\alpha_i$.
The Chevalley invariants of ${\cal G}$  provide for the constants of motion.

The first important result in the search for integrable cases of system (\ref{a1}) is due to Adler and van Moerbeke
\cite{avm2}. They considered the special case where the number of elements in the spectrum $\Delta$ is $n+1$ (i.e.,  $N=n+1$). 
Furthermore, they made the assumption that any $n$ vectors in the spectrum are independent.
Under these
conditions a criterion for algebraic integrability is that

\be
{2 ({\bf v}_i,\, {\bf v}_j )  \over ({\bf v}_i,\, {\bf v}_i ) }  \label{a22}
\ee
should be in the set  $\{ 0, -1, -2, \dots  \}  $.  The classification  obtained corresponds to  the simple roots
of graded Kac--Moody algebras. The associated systems  are the periodic Toda lattices of Bogoyavlensky \cite{bogo3}. The complete 
integrability of these systems using Lax pairs with a spectral parameter was already established in \cite{avm}.
 The method of proof in \cite{avm2} is based on the classical method of Kovalevskaya.

Sklyanin \cite{sklyanin} pointed out  another integrable generalization of the Toda lattice:
\be
  H(q_1, \dots, q_n, \,  p_1, \dots, p_n) = \sum_{i=1}^n \,  { 1 \over 2} \, p_i^2 + 
\sum _{i=1}^{n-1} \,  e^{ q_i-q_{i+1}}  + \alpha_1 e^{q_1}+ \beta_1 e^{2 q_1} + \alpha_n e^{-q_n}+\beta_n e^{-2 q_n}
 \ .  \label{a4}
\ee
He obtained this system  by means of the quantum inverse scattering R-matrix method. 
This potential will be the main focus of this paper.

The next development in the study of system (\ref{a1}) is the work of  Kozlov and Treshchev on  Birkhoff integrable systems.  A system of the form (\ref{a1}) is called
Birkhoff integrable if it has $n$ integrals, polynomial in the momenta with coefficients of the form 
\bd
\sum f_j e^{ <{\bf c}_j,\  {\bf q}>}\ ,   \ \ \ f_j \in {\bf R}, \ \ \  {\bf c}_j \in {\bf R}^n   \ ,
\ed
whose leading homogeneous forms are almost everywhere independent. We remark that in the definition given in the book of Kozlov \cite{kozlov3} there is no
assumption on  involutivity of the integrals. In \cite{kozlov} it is proved that the polynomial integrals are in involution. The terminology has its origin in the work of 
Birkhoff who studied the conditions for the existence of linear and quadratic integrals  of general
Hamiltonians in  two degrees of freedom.  A vector in $\Delta$  is called maximal if it has the greatest possible length among all the vectors in the 
spectrum having the same direction. Kozlov and Treshchev proved the following theorem:

\begin{theorem}
Assume that the Hamiltonian (\ref{a1}) is  Birkhoff integrable. Let ${\bf v}_i$ be a maximal vector in $\Delta$
and assume that the vector ${\bf v}_j \in \Delta$ is linearly independent of ${\bf v}_i$ Then
\bd
{2 ({\bf v}_i,\, {\bf v}_j )  \over ( {\bf v}_i,\, {\bf v}_i ) } 
\ed
lies  in the set $\{ 0, -1, -2, \dots  \}  $.

\end{theorem}

Note that the condition of the theorem is exactly the same as condition (\ref{a22}) of Adler and van Moerbeke. Of course theorem 1 is more general since there is no restriction
on the integer $N$ (the number of summands in the potential of (\ref{a1})). It turns out, however,  that $N$ cannot be much bigger than $n$. In fact, it follows from the classification
that $N \le n+3$.
 A system of the form (\ref{a1}) is called complete if there exist no vector ${\bf v}$ such that the set
$\Delta \cup \{ {\bf v} \}$ satisfies the assumptions of theorem 1.
In \cite{kozlov} there is  a complete classification of all possible  Birkhoff integrable systems based on 
theorem 1. The Dynkin type diagram of a 
complete, irreducible, Birkhoff integrable Hamiltonian system is  isomorphic to one of the following diagrams:

$\put(65,55){\circle{10}}
\put(20,10){\circle{10}}
\put(50,10){\circle{10}}
\put(80,10){\circle{10}}
\put(110,10){\circle{10}}
\put(65,65){1}
\put(20,20){1}
\put(50,20){1}
\put(80,20){1}
\put(110,20){1}
\put(25,10){\line(1,0){20}}
\put(56.5,10){.....}
\put(85,10){\line(1,0){20}}
\put(23.5,13.5){\line(1,1){37.9}}
\put(106.4,13.5){\line(-1,1){37.9}}
\put(60,-20){(a)}$
$\put(190,60){\circle{10}}
\put(280,60){\circle{10}}
\put(190,30){\circle{10}}
\put(190,0){\circle{10}}
\put(220,30){\circle{10}}
\put(250,30){\circle{10}}
\put(280,30){\circle{10}}
\put(280,0){\circle{10}}
\put(195,65){1}
\put(285,65){1}
\put(195,35){1}
\put(195,5){1}
\put(225,35){1}
\put(255,35){1}
\put(285,35){1}
\put(285,5){1}
\put(195,30){\line(1,0){20}}
\put(255,30){\line(1,0){20}}
\put(226.5,30){.....}
\put(190,55){\line(0,-1){20}}
\put(190,25){\line(0,-1){20}}
\put(280,55){\line(0,-1){20}}
\put(280,25){\line(0,-1){20}}
\put(230,-20){(b)}$
$\put(340,0){\circle{10}}
\put(370,0){\circle{10}}
\put(400,0){\circle{10}}
\put(430,0){\circle{10}}
\put(460,0){\circle{10}}
\put(400,30){\circle{10}}
\put(400,60){\circle{10}}
\put(345,5){1}
\put(375,5){1}
\put(405,5){1}
\put(435,5){1}
\put(465,5){1}
\put(405,35){1}
\put(405,65){1}
\put(345,0){\line(1,0){20}}
\put(375,0){\line(1,0){20}}
\put(405,0){\line(1,0){20}}
\put(435,0){\line(1,0){20}}
\put(400,55){\line(0,-1){20}}
\put(398,11){\vdots}
\put(395,-20){(c)}$
$\put(20,-90){\circle{10}}
\put(50,-90){\circle{10}}
\put(80,-90){\circle{10}}
\put(110,-90){\circle{10}}
\put(140,-90){\circle{10}}
\put(170,-90){\circle{10}}
\put(200,-90){\circle{10}}
\put(110,-60){\circle{10}}
\put(25,-85){1}
\put(55,-85){1}
\put(85,-85){1}
\put(115,-85){1}
\put(145,-85){1}
\put(175,-85){1}
\put(205,-85){1}
\put(115,-55){1}
\put(25,-90){\line(1,0){20}}
\put(55,-90){\line(1,0){20}}
\put(85,-90){\line(1,0){20}}
\put(115,-90){\line(1,0){20}}
\put(145,-90){\line(1,0){20}}
\put(175,-90){\line(1,0){20}}
\put(110,-65){\line(0,-1){20}}
\put(100,-120){(d)}$
$\put(250,-90){\circle{10}}
\put(280,-90){\circle{10}}
\put(310,-90){\circle{10}}
\put(340,-90){\circle{10}}
\put(370,-90){\circle{10}}
\put(400,-90){\circle{10}}
\put(430,-90){\circle{10}}
\put(460,-90){\circle{10}}
\put(400,-60){\circle{10}}
\put(255,-85){1}
\put(285,-85){1}
\put(315,-85){1}
\put(345,-85){1}
\put(375,-85){1}
\put(405,-85){1}
\put(435,-85){1}
\put(465,-85){1}
\put(405,-55){1}
\put(255,-90){\line(1,0){20}}
\put(285,-90){\line(1,0){20}}
\put(315,-90){\line(1,0){20}}
\put(345,-90){\line(1,0){20}}
\put(375,-90){\line(1,0){20}}
\put(405,-90){\line(1,0){20}}
\put(435,-90){\line(1,0){20}}
\put(400,-65){\line(0,-1){20}}
\put(325,-120){(e)}$
$\put(20,-160){\circle{10}}
\put(50,-160){\circle{10}}
\put(80,-160){\circle{10}}
\put(110,-160){\circle{10}}
\put(140,-160){\circle{10}}
\put(25,-155){1}
\put(55,-155){1}
\put(85,-155){1}
\put(115,-155){2}
\put(145,-155){2}
\put(25,-160){\line(1,0){20}}
\put(55,-160){\line(1,0){20}}
\put(115,-160){\line(1,0){20}}
\put(84.5,-158.5){\line(1,0){21}}
\put(84.5,-161.5){\line(1,0){21}}
\put(100,-190){(f)}$
$\put(270,-160){\circle{10}}
\put(300,-160){\circle{10}}
\put(330,-160){\circle{10}}
\put(360,-160){\circle{10}}
\put(390,-160){\circle{10}}
\put(275,-155){2}
\put(305,-155){2}
\put(335,-155){2}
\put(365,-155){1}
\put(395,-155){1}
\put(275,-160){\line(1,0){20}}
\put(305,-160){\line(1,0){20}}
\put(365,-160){\line(1,0){20}}
\put(334.5,-158.5){\line(1,0){21}}
\put(334.5,-161.5){\line(1,0){21}}
\put(325,-190){(g)}$
$
\put(20,-240){\circle{10}}
\put(20,-280){\circle{10}}
\put(70,-260){\circle{10}}
\put(100,-260){\circle{10}}
\put(130,-260){\circle{10}}
\put(160,-260){\circle{10}}
\put(210,-240){\circle{10}}
\put(210,-280){\circle{10}}
\put(25,-235){4}
\put(30,-270){1}
\put(75,-255){2}
\put(105,-255){2}
\put(135,-255){2}
\put(170,-265){2}
\put(215,-235){4}
\put(215,-275){1}
\put(75,-260.5){\line(1,0){20}}
\put(135,-260){\line(1,0){20}}
\put(106.5,-260){.....}
\put(18.3,-245){\line(0,-1){30.5}}
\put(21.6,-245){\line(0,-1){30.5}}
\put(208.3,-245){\line(0,-1){30.5}}
\put(211.6,-245){\line(0,-1){30.5}}
\put(24,-244){\line(5,-2){41}}
\put(24,-237){\line(5,-2){45}}
\put(24,-284){\line(5,2){46.5}}
\put(24,-277){\line(5,2){41}}
\put(206,-244){\line(-5,-2){41}}
\put(210,-235){\line(-5,-2){48.5}}
\put(206,-284){\line(-5,2){46.5}}
\put(206,-277){\line(-5,2){41}}
\put(15,-240){\line(0,-1){40}}
\put(24,-244){\line(0,-1){34}}
\put(215,-240){\line(0,-1){40}}
\put(206,-244){\line(0,-1){34}}
\put(100,-310){(h)}$
$
\put(450,-260){\circle{10}}
\put(350,-260){\circle{10}}
\put(400,-240){\circle{10}}
\put(400,-280){\circle{10}}
\put(455,-255){2}
\put(345,-255){2}
\put(405,-235){4}
\put(405,-295){1}
\put(398.3,-245){\line(0,-1){30.5}}
\put(401.6,-245){\line(0,-1){30.5}}
\put(404,-244){\line(5,-2){41}}
\put(404,-237){\line(5,-2){45}}
\put(404,-284){\line(5,2){46.5}}
\put(404,-277){\line(5,2){41}}
\put(396,-244){\line(-5,-2){41}}
\put(400,-235){\line(-5,-2){48.5}}
\put(396,-284){\line(-5,2){46.5}}
\put(396,-277){\line(-5,2){41}}
\put(405,-240){\line(0,-1){40}}
\put(396,-244){\line(0,-1){34}}
\put(390,-310){(i)}$
$
\put(20,-380){\circle{10}}
\put(20,-410){\circle{10}}
\put(20,-440){\circle{10}}
\put(50,-410){\circle{10}}
\put(80,-410){\circle{10}}
\put(110,-410){\circle{10}}
\put(160,-390){\circle{10}}
\put(160,-430){\circle{10}}
\put(25,-375){2}
\put(25,-435){2}
\put(25,-405){2}
\put(55,-405){2}
\put(85,-405){2}
\put(120,-415){2}
\put(165,-385){4}
\put(165,-425){1}
\put(20,-385){\line(0,-1){20}}
\put(20,-415){\line(0,-1){20}}
\put(25,-410.5){\line(1,0){20}}
\put(85,-410){\line(1,0){20}}
\put(56.5,-410){.....}
\put(158.3,-395){\line(0,-1){30.5}}
\put(161.6,-395){\line(0,-1){30.5}}
\put(156,-394){\line(-5,-2){41}}
\put(160,-385){\line(-5,-2){48.5}}
\put(156,-434){\line(-5,2){46.5}}
\put(156,-427){\line(-5,2){41}}
\put(165,-390){\line(0,-1){40}}
\put(156,-394){\line(0,-1){34}}
\put(100,-470){(j)}$
$\put(410,-410){\circle{10}}
\put(410,-380){\circle{10}}
\put(410,-440){\circle{10}}
\put(330,-410){\circle{10}}
\put(360,-410){\circle{10}}
\put(417,-410){4}
\put(415,-375){9}
\put(415,-452){1}
\put(330,-403){3}
\put(360,-403){3}
\put(405,-380){\line(0,-1){60}}
\put(415,-380){\line(0,-1){60}}
\put(408.3,-384.5){\line(0,-1){21}}
\put(411.6,-384.5){\line(0,-1){21}}
\put(408.3,-414.5){\line(0,-1){21}}
\put(411.6,-414.5){\line(0,-1){21}}
\put(335,-410){\line(1,0){20}}
\put(365,-410){\line(1,0){40}}
\put(360,-405){\line(1,0){50}}
\put(360,-415){\line(1,0){50}}
\put(359.5,-404){\line(5,3){49}}
\put(365,-410){\line(5,3){43}}
\put(363.25,-406){\line(5,3){41.5}}
\put(359.5,-416){\line(5,-3){49}}
\put(365,-410){\line(5,-3){43}}
\put(363.25,-414){\line(5,-3){41.5}}
\qbezier(410,-385)(465,-410)(410,-435)
\qbezier(413,-381.7)(470,-410)(413,-438.4)
\qbezier(413,-378.4)(475,-410)(413,-441.7)
\qbezier(410,-375)(485,-410)(410,-445)
\put(390,-470){(k)}
$

\vskip 1cm

\noindent
{\it Remark 1}
In the list of  diagrams we have omitted some cases that occur as sub--graphs of diagrams (a)--(k) (by truncating
one or more vertices).  In other words, the spectrum of a Birkhoff integrable Hamiltonian system is obtained  from the spectrum of a complete system
by dropping part of the elements.

\noindent
{\it Remark 2} 
The Dynkin type diagram determines only the angles between pairs of vectors in $\Delta$. In order to reconstruct the ratios of lengths of vectors in $\Delta$ we assign to
the $i$th vertex a coefficient proportional to the square of the length  of ${\bf v}_i$.  This explains the numbers appearing on  the vertices of the diagrams.

We have to stress that this classification gives only necessary conditions for a system of type (\ref{a1}) to
be Birkhoff integrable. The integrability for each system in the list should be established case by case. 
We give a brief history of the progress in this direction.
As we already mentioned,  the integrability of systems (a)--(g) was established in \cite{avm},  \cite{bogo3}. 
The  solution of these generalized periodic Toda lattices (associated with affine Lie algebras)  was obtained by Goodman and Wallach  in \cite{goodman}. 
The graph  (i) corresponds to a Hamiltonian system in  two
degrees of freedom with potential
\bd
e^{q_1}+e^{q_2}+e^{-q_1-q_2}+e^{ -\left( {q_1+ q_2 \over 2} \right)} \ .
\ed
The additional integral can be found in \cite{kozlov}.  The integrability or non--integrability
 of systems (j) and (k) is still open. No Lax pair is known for either system.  It is believed that system (j) is completely integrable, in fact 
integrability is established in \cite{emelyanov} for the case $n=4$. It is generally believed that system
(k) in non--integrable.

In this paper we deal with the integrability of  system $(h)$ which corresponds to the Hamiltonian
(\ref{a4}). Sklyanin in \cite{sklyanin} indicated this
system as another integrable generalization of the Toda lattice. The case $n=2$ corresponds to the potential
\bd
V=e^{q_1-q_2}+ c_1 e^{2 q_2}+c_2 e^{q_2}+ c_3^{-q_1} + c_4 e^{-2 q_1} \ .
\ed
Annamalai and Tamizhmani  \cite{anna} demonstrated  the integrability of this particular case  by using Noether's theorem.
The second integral is of fourth degree in the momenta.

The case $n=3$ (as well as the general case)  is  treated in Ranada \cite{ranada}. Ranada  proved integrability by
using  a Lax pair approach. The additional integrals are of degree 4 and 6. 

In this paper we examine  the integrability of the system

\be
  H(q_1, \dots, q_n, \,  p_1, \dots, p_n) = \sum_{i=1}^n \,  { 1 \over 2} \, p_i^2 + 
\sum _{i=1}^{n-1} \,  e^{ q_i-q_{i+1}}  + \alpha e^{-2 q_1}+ \beta e^{2 q_n} \label{a6} \ .
\ee
Without loss of generality we may assume that $\alpha=\beta=1$.  We note that if either $\alpha$ or $\beta$ are zero then the
system reduces to well-known cases of generalized Toda lattices..
The technique we use is entirely new and hopefully it will lead to further results for  similar systems.
The strategy is the following:
 We consider the $D_n$  Volterra system whose Lax pair formulation and multi--Hamiltonian
structure was established recently in Kouzaris \cite{kouzaris}.
Using a procedure  of Moser we transform the system into a new system in some intermediate variables $(a_i, b_i)$:

\begin{eqnarray}
\dot{a}_{1} &=&2a_{1}b_{1} \label{a7} \\
\dot{a}_{i} &=&a_{i}\left( b_{i}-b_{i-1}\right) \ \ \ \ i=2,3,\ldots , n \nonumber \\
\dot{a}_{n+1} &=&-2a_{n+1}b_{n} \nonumber \\
\dot{b}_{i} &=&2\left( a_{i+1}^{2}-a_{i}^{2}\right) \ \ \  \  i=1,2,\ldots ,n     \ .  \nonumber 
\end{eqnarray}

We obtain a Lax pair for this system and the integrals of motion. The final step is the construction of a 
Flaschka  transformation from the Hamiltonian system (\ref{a6}) to the system  (\ref{a7}). The inverse of Flaschka's
transformation provides for the necessary constants of motion in the variables $(q,p)$ for the system (\ref{a6}).
Thus, integrability is established.

In section 2 we describe the construction of Bogoyavlensky--Volterra systems following \cite{bogo1}, \cite{bogo2}.
Bogoyavlensky constructed the systems using the root system of a simple Lie algebra and then through a change
of variables (from $c_i$ to $u_i$) he ended--up with  homogeneous polynomial systems in the new variables
$u_i$. The construction of Lax pairs in the  variables $u_i$ is  in \cite{kouzaris}. 

In section 3 we present  part of the results of \cite{kouzaris}, namely the $D_n$ case, since it is the
only system needed for  the purposes of the present paper. We have to point out that in  \cite{kouzaris} there is
a complete treatment of the multi--Hamiltonian structure, Lax pairs and integrability
of Bogoyavlensky--Volterra lattices.

The results of section 4 are entirely new.  They  can be summarized as  two transformations, one Moser--type
from the Volterra $D_n$ lattice  to  system (\ref{a7}) and one Flaschka--type from the Sklyanin system (\ref{a6}) to
system (\ref{a7}). Since Flaschka--type transformations are well--known, we describe briefly the Moser approach.

Consider the system
\be
\frac{du_{i}}{dt}=u_{i}(u_{i+1}-u_{i-1})\qquad i=1,...,n\ , \label{a33}
\ee
where $u_{0}=u_{n+1}=0.$ This is the Volterra system, also known as the KM
system and\ is related to the root system of a simple Lie algebra of type $A_{n}.$
The infinite KM--system was solved by Kac and Van
Moerbeke \cite{kac} using a discrete version of inverse scattering. 
The Lax pair for system (\ref{a33}) can be found in  \cite{moser2}. The Lax matrix has the form
\be
L=\begin{pmatrix}{ 0 & a_1& 0 & 0& \dots & \cr
                 a_1&0& a_2& 0& \dots & \cr
                 0 & a_2 & a_3 & 0 & \cdots & \cr
                 \vdots & \vdots & \vdots& & & \cr
                  & && & & \cr
                  &&&&& a_{n-1} \cr
                  &&&&a_{n-1} & 0} \end{pmatrix} \ ,
\ee
where  $u_i=2 a_i^2$.
Moser in \cite{moser2} describes a relation between the KM system  (\ref{a33}) and the non--periodic Toda lattice. The procedure is
the following: Form $L^2$  which is not anymore a tridiagonal matrix but is similar to one.  Let $\{e_1, e_2, \dots, e_n \}$ be the standard
basis of ${\bf R}^n$, and $E_o= \{ {\rm span}\, e_{2i-1}, \,  i=1,2, \dots \}$, $E_e= \{ {\rm span}\, e_{2i}, \, i=1,2, \dots \}$. Then $L^2$ leaves
$E_o$,  $E_e$ invariant and reduces to each of these spaces to a tri--diagonal symmetric Jacobi matrix.
 For example, if we omit  all even columns and all even  rows we 
obtain a tridiagonal Jacobi matrix and the entries of this new matrix  define the transformation from the KM--system
to the Toda lattice. We illustrate with a simple example where $n=5$.

In this case 
\be
L=\begin{pmatrix}{ 0 & a_1& 0 & 0& 0\cr
                 a_1&0& a_2& 0& 0\cr
                 0 & a_2 &0 & a_3 & 0  \cr
                 0 & 0 & a_3& 0 &a_4 \cr
                  0 &0&0&a_4& 0 } \ ,
                  \end{pmatrix}
\ee

and $L^2$ is the matrix

\be
\begin{pmatrix}{ a_1^2 & 0& a_1 a_2 & 0& 0 \cr
                 0& a_1^2+a_2^2& 0& a_2 a_3&  0\cr
                 a_1 a_2 & 0 & a_2^2+a_3^2 & 0 & a_3 a_4\cr
                 0 & a_2 a_3 & 0& a_3^2+a_4^2 &0\cr
                  0 &0& a_3 a_4&0&a_4^2 } \ .
                  \end{pmatrix}
\ee
Omitting even  columns and even rows of $L^2$ we obtain the matrix
\be
\begin{pmatrix} { a_1^2 & a_1 a_2 & 0 \cr
                   a_1 a_2 &a_2^2+a_3^2 & a_3 a_4 \cr
                    0 & a_3 a_4 & a_4^2 } \ .
                  \end{pmatrix}
\ee
This is a tridiagonal Jacobi matrix. It is natural to define   new variables $A_1=a_1 a_2$, $A_2=a_3 a_4$, $B_1=a_1^2$, $B_2=a_2^2+a_3^2$, $B_3=a_4^2$. The new 
 variables $A_1,A_2, B_1,B_2, B_3$ satisfy the Toda lattice  equations.

This procedure shows that the KM-system  and the Toda lattice are closely related: The explicit  transformation
 which is  due to H\'enon 
maps one system to the other. The mapping in the general case  is given by 
\be
A_i=-{ 1 \over 2} \sqrt {u_{2i} u_{2i-1}} \  , \qquad  B_i= { 1 \over 2}\left( u_{2i-1}+u_{2i-2} \right)  \label{a25} \ .
\ee
The equations satisfied by the new variables $A_i$, $B_i$ are given by:

\begin{displaymath} 
\begin{array}{lcl}
 \dot A _i& = & A_i \,  (B_{i+1} -B_i )    \\  
   \dot B _i &= & 2 \, ( A_i^2 - A_{i-1}^2 ) \ . 
\end{array}
\end{displaymath}    
These are precisely the Toda equations in \cite{flaschka1}.
We remark that the transformation (\ref{a25}) was first discovered by H\'enon. H\'enon never published the result but he communicated the formula  in
a letter to Flaschka in 1973. We refer to paper
\cite{damianou3} for more details. In \cite{damianou3} one can find the multiple--Hamiltonian structure, higher Poisson structures and 
master symmetries for system (\ref{a33}). 

We would like to   generalize the H\'enon correspondence (\ref{a25})  (using the recipe of Moser) from generalized Volterra to generalized Toda systems.
The relation between the Volterra systems of type $B_n$ and $C_n$ and the corresponding Toda systems is in \cite{damianou1}. 
It is natural to attempt to find a similar correspondence between the Volterra lattice   of type $D_n$ and the generalized
Toda lattice of type $D_n$. It is a surprising result, and this is the content of the present paper that the Volterra $D_n$ system
corresponds not to the Toda $D_n$ system but to a special case of the Sklyanin lattice.

\section{Bogoyavlensky--Volterra systems}

Bogoyavlensky constructed integrable Hamiltonian  systems connected with simple
Lie algebras,  generalizing    the KM--system (\ref{a33}). For more details
see ref. \cite{bogo1}, \cite{bogo2}. In  this section we summarize   the construction of Bogoyavlensky.

Let\textit{\ }$\mathcal{G}$ be a simple Lie algebra of rank $n$
and $\Pi =\{\omega _{1,}\, \omega _{2},\ldots ,\, \omega _{n}\}$ the Cartan-Weyl
basis of the simple roots in $\mathcal{G}$. There exist
unique, positive integers $k_{i}$ such that%
\bd
k_{0}\, \omega _{0}+k_{1}\, \omega _{1}+\cdots +k_{n}\, \omega _{n}=0 \ ,
\ed
where $k_{0}=1$ and $\omega _{0}$ is the minimal negative root.

We consider the following Lax pairs: 
\begin{eqnarray}
\dot{L} &=&\left[ B,L\right] \ , \label{a8}  \\
L\left( t\right) &=&\sum_{i=1}^{n}c_{i}\left( t\right) e_{\omega
_{i}}+e_{\omega _{0}}+\sum_{1\leq i<j\leq n}\left[ e_{\omega
_{i}},e_{\omega _{j}}\right]\ ,  \nonumber \\
B\left( t\right) &=&\sum_{i=1}^{n}\frac{k_{i}}{c_{i}\left( t\right) }%
e_{-\omega _{i}}+e_{-\omega _{0}}\ .  \nonumber
\end{eqnarray}

Let $\mathcal{H}$ be a Cartan subalgebra of $\mathcal{G}$ . For every root $%
\omega _{a}\in $ $\mathcal{H}^{\ast }$ there is a unique $H_{\omega _{a}}\in 
\mathcal{H}$ such that $\omega \left( h\right) =k\left( H_{\omega
_{a}},h\right) $ $\forall $ $h\in \mathcal{H,}$ where $k$ is the Killing
form. We also
have an inner product on $\mathcal{H}^{\ast }$ such that $\left\langle \omega
_{a},\omega _{b}\right\rangle =k\left( H_{\omega _{a}},H_{\omega
_{b}}\right) $. We set

\bd
c_{ij}=\left\{ 
\begin{array}{cc}
1 & { \rm if} \ \ \left\langle \omega _{i},\omega _{j}\right\rangle \neq 0 \ \  {\rm
and } \ \ \ i<j \\ 
0 & {\rm if } \  \ \left\langle \omega _{i},\omega _{j}\right\rangle =0 \ \ \  {\rm or } \ \ \
i=j    \\ 
-1 \  & {\rm if } \ \ \left\langle \omega _{i},\omega _{j}\right\rangle
\neq 0 \ \ \ { \rm and } \ \ \ i>j  \ .
\end{array}%
\right.
\ed

The matrix  equation (\ref{a8}) is equivalent to the dynamical system

\begin{equation}
\dot{c}_{i}=-\sum_{j=1}^{n}\frac{k_{j}c_{ij}}{c_{j}}\label{a9}\ .
\end{equation}

We determine the skew-symmetric variables 
\bd
\begin{array}{ccc}
x_{ij}=c_{ij}c_{i}^{-1}c_{j}^{-1}, & x_{ji}=-x_{ij}, & x_{jj}=0 \ ,
\end{array}%
\ed
which correspond to the edges of the Dynkin diagram for the Lie algebra $%
\mathcal{G}$,  connecting the vertices $\omega _{i}$ and $\omega _{j}.$

The dynamical system (\ref{a9})  in the variables $x_{ij}$ takes the
form%
\begin{equation}
\dot{x}_{ij}=x_{ij}\sum_{s=1}^{n}k_{s}\left( x_{is}+x_{js}\right)  \label{a10} \ .
\end{equation}

We recall that the vertices $\omega _{i},\ \omega _{j}$ of the Dynkin diagram
are joined by edges only if $\left\langle \omega _{i},\omega
_{j}\right\rangle \neq 0$. Hence $x_{ij}=0$ if there are no edges
connecting the vertices $\omega _{i}$ and $\omega _{j}$ of the diagram. We
call the equations (\ref{a10}) the  Bogoyavlensky-Volterra system
associated with $\mathcal{G}$ ($BV$ system for short).

We shall now describe the $BV$ system for each simple Lie algebra $\mathcal{G%
}$. The number of independent variables $x_{ij}\left( t\right) $ is equal to 
$n-1$ and is one less than the number of variables $c_{j}\left( t\right) $ .
We use the standard numeration of vertices of the Dynkin diagram and define
the variables $u_{k}\left( t\right) =x_{ij}\left( t\right) $ corresponding
to the edges of the Dynkin diagram with increasing order of the vertices $%
\left( i<j\right) .$

The phase space consists of variables $u_i$,  with $u_i>0$.
In the following list we give explicit expressions for the Volterra  systems in 
the variables $u_i$ for each classical simple Lie algebra.
 \newpage

\newpage
\centerline{\Large {\bf $A_{n+1}$}}
\vskip 1cm

$\bigskip 
\put(25,20){\line(1,0){50}}
\put(85,20){\line(1,0){50}}
\put(205,20){\line(1,0){50}}
\put(20,20){\circle{10}}
\put(80,20){\circle{10}}
\put(140,20){\circle{10}}
\put(200,20){\circle{10}}
\put(260,20){\circle{10}}
\put(17,30){$\omega_1$}
\put(77,30){$\omega_2$}
\put(137,30){$\omega_3$}
\put(197,30){$\omega_{n}$}
\put(257,30){$\omega_{n+1}$}
\put(45,13){$u_1$}
\put(105,13){$u_2$}
\put(225,13){$u_n$}
\put(160,20){$\ldots$}
$

\bigskip $%
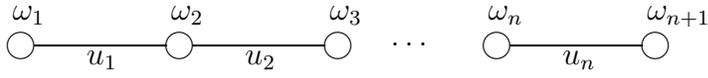
\begin{tabular}{ll}
\begin{tabular}{l}
$\omega _{0}=-\left( \omega _{1}+\omega _{2}+\cdots \omega _{n+1}\right) $
\\ 
$k_{i}=1$ , $\ i=1,\ldots ,n+1$ \\ 
\\ 
$c_{ij}=\left\{ 
\begin{array}{c}
\ \ \ \ 0\ \  \ \ \left| i-j\right| \neq 1 \\ 
\begin{tabular}{ll}
$\ \ \ 1$ &  \ \  $j=i+1$%
\end{tabular}
\\ 
\begin{tabular}{ll}
$-1$ &  \ \  $j=i-1$%
\end{tabular}%
\end{array}%
\right. $ \\ 
\\ 
$u_{i}=x_{i,i+1}=\frac{1}{c_{i}c_{i+1}}$ \  \ \ \ $i=1,\ldots ,n$%
\end{tabular}
& \ \ \ \ \ \ \ \ \ \ \ $\bf{\ \ }%
\begin{tabular}{l}

\\ 
$\dot{u}_{1}$ $\ =u_{1}u_{2}$ \\ 
$\dot{u}_{n}$ $=-u_{n-1}u_{n}$ \\ 
$\dot{u}_{i}$ $\ =u_{i}(u_{i+1}-u_{i-1})$ \\ 
$2\leq i\leq n-1$%
\end{tabular}%
$\ 
\end{tabular}%
$

$%
\newline
\put(25,0){\line(1,0){250}}
\newline%
$

\vskip 1cm
\centerline{\Large {\bf $B_{n+1}$}}
\vskip 1cm

$%
\put(25,20){\line(1,0){50}}
\put(85,20){\line(1,0){50}}
\put(205,19){\line(1,0){50}}
\put(205,21){\line(1,0){50}}
\put(20,20){\circle{10}}
\put(80,20){\circle{10}}
\put(140,20){\circle{10}}
\put(200,20){\circle{10}}
\put(260,20){\circle{10}}
\put(17,30){$\omega_1$}
\put(77,30){$\omega_2$}
\put(137,30){$\omega_3$}
\put(197,30){$\omega_{n}$}
\put(257,30){$\omega_{n+1}$}
\put(45,13){$u_1$}
\put(105,13){$u_2$}
\put(218,13){$u_n$}
\put(160,20){$\ldots$}
\put(227,17.5){$\gg $}%
$

\bigskip $%
\begin{tabular}{ll}
\begin{tabular}{l}
$\omega _{0}=-\left( \omega _{1}+2\omega _{2}+\cdots +2\omega _{n+1}\right) $
\\ 
$k_{1}=1, \ \ k_{i}=2$ , $\ i=2,\ldots ,n+1$ \\ 
\\ 
$c_{ij}=\left\{ 
\begin{array}{c}
\ \ \ \ 
\begin{tabular}{ll}
$0$ &  \ \ $\ \left| i-j\right| \neq 1$%
\end{tabular}
\\ 
\begin{tabular}{ll}
$\ \ \ 1$ & \ \  $j=i+1$%
\end{tabular}
\\ 
\begin{tabular}{ll}
$-1$ &  \ \  $j=i-1$%
\end{tabular}%
\end{array}%
\right. $ \\ 
\\ 
$u_{i}=x_{i,i+1}=\frac{1}{c_{i}c_{i+1}}$ \ \ \ \ $i=1,\ldots ,n$%
\end{tabular}
& \ \ \ \ \ \ \ \ \ \ \ $\bf{\ \ }%
\begin{tabular}{l}

\\ 
$\dot{u}_{1}$ $\ =u_{1}\left( u_{1}+2u_{2}\right) $ \\ 
$\dot{u}_{2}$ $\ =u_{2}\left( 2u_{3}-u_{1}\right) $ \\ 
$\dot{u}_{n}$ $=-2u_{n-1}u_{n}$ \\ 
$\dot{u}_{i}$ $\ =2u_{i}(u_{i+1}-u_{i-1})$ \\ 
$3\leq i\leq n-1$%
\end{tabular}%
$\ 
\end{tabular}%
$

\newpage
\centerline{\Large {\bf $C_{n+1}$}}
\vskip 1cm

$%
\put(25,20){\line(1,0){50}}
\put(85,20){\line(1,0){50}}
\put(205,19){\line(1,0){50}}
\put(205,21){\line(1,0){50}}
\put(20,20){\circle{10}}
\put(80,20){\circle{10}}
\put(140,20){\circle{10}}
\put(200,20){\circle{10}}
\put(260,20){\circle{10}}
\put(17,30){$\omega_1$}
\put(77,30){$\omega_2$}
\put(137,30){$\omega_3$}
\put(197,30){$\omega_{n}$}
\put(257,30){$\omega_{n+1}$}
\put(45,13){$u_1$}
\put(105,13){$u_2$}
\put(218,13){$u_n$}
\put(160,20){$\ldots$}
\put(227,17.5){$\ll $}%
$

\bigskip $%
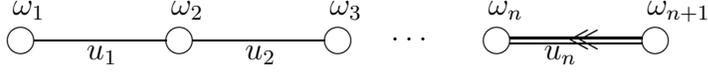
\begin{tabular}{ll}
\begin{tabular}{l}
$\omega _{0}=-\left( 2\omega _{1}+\cdots +2\omega _{n}+\omega _{n+1}\right) $
\\ 
$k_{i}=2$ $,\ i=1,\ldots ,n$ $, \ k_{n+1}=1$ \\ 
\\ 
$c_{ij}=\left\{ 
\begin{array}{c}
\ \ \ \ 
\begin{tabular}{ll}
$0$ & \ \ $\ \left| i-j\right| \neq 1$%
\end{tabular}
\\ 
\begin{tabular}{ll}
$\ \ \ 1$ & \ \  $j=i+1$%
\end{tabular}
\\ 
\begin{tabular}{ll}
$-1$ & \ \  $j=i-1$%
\end{tabular}%
\end{array}%
\right. $ \\ 
\\ 
$u_{i}=x_{i,i+1}=\frac{1}{c_{i}c_{i+1}}$ \  \ $i=1,\ldots ,n$%
\end{tabular}
& \ \ \ \ \ \ \ $\bf{\ }%
\begin{tabular}{l}

\\ 
$\dot{u}_{1}$ $\ \ \ =2u_{1}u_{2}$ \\ 
$\dot{u}_{n-1}$ $=u_{n-1}(u_{n}-2u_{n-2})$ \\ 
$\dot{u}_{n}$ $\ \ \ =-u_{n}(u_{n}+2u_{n-1})$ \\ 
$\dot{u}_{i}$ $\ \ \ =2u_{i}(u_{i+1}-u_{i-1})$ \\ 
$2\leq i\leq n-2$%
\end{tabular}%
$\ \ \ \ $\bf{\ \ }$%
\end{tabular}%
$

$%
\newline
\put(25,0){\line(1,0){250}}
\newline%
$

\vskip 1cm
\centerline{\Large {\bf $D_{n+1}$}}
\vskip 1cm

$
\put(25,20){\line(1,0){50}}
\put(85,20){\line(1,0){50}}
\put(205,20){\line(1,0){50}}
\put(263.5,23.5){\line(1,1){33}}
\put(263.5,16.5){\line(1,-1){33}}
\put(20,20){\circle{10}}
\put(80,20){\circle{10}}
\put(140,20){\circle{10}}
\put(200,20){\circle{10}}
\put(260,20){\circle{10}}
\put(300,60){\circle{10}}
\put(300,-20){\circle{10}}
\put(17,30){$\omega_1$}
\put(77,30){$\omega_2$}
\put(137,30){$\omega_3$}
\put(197,30){$\omega_{n-2}$}
\put(270,20){$\omega_{n-1}$}
\put(45,13){$u_1$}
\put(105,13){$u_2$}
\put(222,13){$u_{n-2}$}
\put(160,20){$\ldots$}
\put(297,47){$\omega_n$}
\put(297,-8){$\omega_{n+1}$}
\put(254,40){$u_{n-1}$}
\put(262,0){$u_{n}$}%
$

$%
\begin{tabular}{l}
$\omega _{0}=-\left( \omega _{1}+2\omega _{2}+\cdots +2\omega _{n-1}+\omega
_{n}+\omega _{n+1}\right) $ \\ 
$k_{1}=1$ $,$ $k_{n}=1$ , $k_{n+1}=1$ $\ ,$ $k_{i}=2$ $,$ $2\leq i\leq n-1$
\\ 
$c_{ij}=-c_{ji}=\left\{ 
\begin{tabular}{ll}
1 & \ \ $2\leq j=i+1\leq n$ \\ 
0 & \ \  $\left( i,j\right) =\left( n,n+1\right) $ \\ 
0 & \ \  $3\leq $ $i+2\leq j\leq n$ \\ 
1 & \ \  $\left( i,j\right) =\left( n-1,n+1\right) $%
\end{tabular}%
\right. $ \\ 
\\ 
$u_{i}=x_{i,i+1}=\frac{1}{c_{i}c_{i+1}}$ , $i=1,\ldots ,n-1$ , $\
u_{n}=x_{n-1,n+1}=\frac{1}{c_{n-1}c_{n+1}}$%
\end{tabular}%
$

\bigskip 
\begin{eqnarray}
\dot{u}_{1} &=&u_{1}\left( 2u_{2}+u_{1}\right) \ , \ \ \ \dot{u}_{2}=u_{2}\left( 2u_{3}-u_{1}\right)\  \nonumber \\
\dot{u}_{i} &=&2 u_{i}(u_{i+1}-u_{i-1}) \  \ 3\leq i\leq n-3 \  \nonumber \\
\dot{u}_{n-2}&=&u_{n-2}\left( u_{n}+u_{n-1}-2u_{n-3}\right) \  \label{a55} \\
\dot{u}_{n-1} &=&u_{n-1}\left( u_{n}-u_{n-1}-2u_{n-2}\right) \  \nonumber \\
\dot{u}_{n}&=&-u_{n}\left( u_{n}-u_{n-1}+2u_{n-2}\right)\ .  \nonumber
\end{eqnarray}

\section{The Volterra $D_n$ system}

Consider the  $BV$ $D_{n+1}$ system (\ref{a55}) in the variables $u_k$.
We make a linear change of variables
\bd
v_{1}=u_{1}, \  v_{k}=2u_{k} \ \ k=2,\ldots n-2, \ 
v_{n-1}=u_{n-1}, \  v_{n}=u_{n} \ ,
\ed
to obtain the equivalent system  

\begin{eqnarray}
\dot{v}_{1}\ &=&v_{1}\left( v_{1}+v_{2}\right) \label{a40}  \\
\dot{v}_{k}\ &=&v_{k}(v_{k+1}-v_{k-1})\ \ k=2,\ldots n-3  \nonumber
\\
\dot{v}_{n-2} &=&v_{n-2}(v_{n}+v_{n-1}-v_{n-3})  \nonumber \\
\dot{v}_{n-1} &=&v_{n-1}(v_{n}-v_{n-1}-v_{n-2})  \nonumber \\
\dot{v}_{n} &=&-v_{n}(v_{n}-v_{n-1}+v_{n-2})\ .  \nonumber
\end{eqnarray}%
\qquad

Before giving the Lax pair for the system (\ref{a40})  we introduce
some matrix notations:%
\begin{eqnarray}
\ X_{k} &=&\left( 
\begin{array}{cc}
\sqrt{v_{k}} & 0 \nonumber \\ 
0 & i\sqrt{v_{k}}%
\end{array}%
\right) \ , \quad  O=\left( 
\begin{array}{cc}
0 & 0 \\ 
0 & 0%
\end{array}%
\right) \ ,  \\
Y_{k} &=&\frac{1}{2}\left( 
\begin{array}{cc}
\sqrt{v_{k}v_{k+1}} & 0 \\ 
0 & \sqrt{v_{k}v_{k+1}}%
\end{array}%
\right) \ , \qquad  Y_{0}=\frac{i}{2}\left( 
\begin{array}{cc}
0 & v_{1} \\ 
-v_{1} & 0%
\end{array}%
\right) \ .  \nonumber
\end{eqnarray}

We also set%
\bd
X=\left( 
\begin{array}{cc}
\sqrt{v_{n}} & i\sqrt{v_{n}}\nonumber \\ 
-\sqrt{v_{n-1}} & i\sqrt{v_{n-1}}%
\end{array}%
\right) \ , \qquad Y=\frac{1}{2}\left( 
\begin{array}{cc}
\sqrt{v_{n-2}v_{n}} & \sqrt{v_{n-2}v_{n}} \\ 
-\sqrt{v_{n-2}v_{n-1}} & \sqrt{v_{n-2}v_{n-1}}%
\end{array}%
\right) ,
\ed
\[
W=\frac{i}{2}\left( 
\begin{array}{cc}
0 & v_{n-1}-v_{n} \\ 
v_{n}-v_{n-1} & 0%
\end{array}%
\right)\ .
\]

Equations (\ref{a40}) can be written in a Lax pair form $\dot{L}=%
\left[ L,B\right] $, where 
\begin{equation}
L=\left[ 
\begin{array}{ccccccc}
0 &  & 0 & \cdots & 0 & \sqrt{v_{1}} & i\sqrt{v_{1}} \\ 
&  &  &  &  &  &  \\ 
0 &  & O & X & O & \cdots & O \\ 
\vdots &  & X^{t} & O & X_{n-2} & \ddots & \vdots \\ 
0 &  & O & X_{n-2} & \ddots & \ddots & O \\ 
\sqrt{v_{1}} &  & \vdots & \ddots & \ddots & O & X_{2} \\ 
i\sqrt{v_{1}} &  & O & \cdots & O & X_{2} & O%
\end{array}%
\right]  \ , \label{a41} 
\end{equation}

\[
B=\left[ 
\begin{array}{ccccccccc}
0 &  & \cdots & \cdots & 0 & -\frac{1}{2}\sqrt{v_{1}v_{2}} & -\frac{1}{2}%
\sqrt{v_{1}v_{2}} & 0 & 0 \\ 
&  &  &  &  &  &  &  &  \\ 
\vdots &  & O & O & Y & O & \cdots & \cdots & O \\ 
\vdots &  & O & W & O & Y_{n-3} & \ddots &  & \vdots \\ 
0 &  & -Y^{t} & O & O & \ddots & \ddots & O & \vdots \\ 
\frac{1}{2}\sqrt{v_{1}v_{2}} &  & O & -Y_{n-3} & \ddots & \ddots & O & Y_{3}
& O \\ 
\frac{1}{2}\sqrt{v_{1}v_{2}} &  & \vdots & \ddots & \ddots & O & O & O & 
Y_{2} \\ 
0 &  & \vdots &  & O & -Y_{3} & O & O & O \\ 
0 &  & O & \cdots & \cdots & O & -Y_{2} & O & Y_{0}%
\end{array}%
\right] \ .
\]
We note  that the entries of the first row and the first column of $L$ and $B$ are scalars
while all the other entries are $2\times 2$ matrices.

The invariant polynomials of this system are given by the functions 
\begin{eqnarray*}
&&H_{2},H_{4},\ldots ,H_{n-1} \ \ {\rm when}\   n \ {\rm is \  odd ,} \\
&&H_{2},H_{4},\ldots ,H_{n-2},H_{n-1} \ \ {\rm when } \ n \ {\rm  is \  even ,}
\end{eqnarray*}%
where $H_{k}=\frac{1}{k}Tr(L^{k})$ .

We use the variables $c_{j},\  \ 1\leq j\leq n+1$ of  equations (\ref{a9})  in order to find a
cubic bracket $\pi _{3}$ of the $BV$ $D_{n+1}$ system. The dynamical system
(\ref{a9})  in the case of the Lie algebra of type $D_{n+1}\ $can be
written in Hamiltonian form $\dot{c}_{j}=\left\{ c_{j},H\right\} $, with
Hamiltonian
 
\bd
H=\log c_{1}+2\sum_{j=2}^{n-1}\log c_{j}+\log c_{n}+\log c_{n+1}\ ,
\ed
and Poisson bracket

\begin{eqnarray}
\left\{ c_{j},c_{j+1}\right\} &=&-\left\{ c_{j+1},c_{j}\right\} =1  \ \ \ j=1,2,\ldots ,n-1  \label{a42} \\
\left\{ c_{n-1}, c_{n+1}\right\} &=&-\left\{ c_{n+1}, c_{n-1}\right\} =1 \ ;
  \nonumber
\end{eqnarray}%
all other brackets are zero. In the new  variables $v_{j}$ $\ \
(v_{1}=c_{1}^{-1}c_{2}^{-1}$ $,$ $v_{k}=2c_{k}^{-1}c_{k+1}^{-1}$ $,$ $%
k=2,\ldots ,n-2,$ $\ \ v_{n-1}=c_{n-1}^{-1}c_{n}^{-1}$ $,$ $%
v_{n}=c_{n-1}^{-1}c_{n+1}^{-1}$ $)$ the above skew-symmetric bracket, which
we denote by $\pi _{3},$ is given by%

\begin{eqnarray}
\left\{ v_{1},v_{2}\right\} &=&v_{1}v_{2}\left( 2v_{1}+v_{2}\right) \label{a43} \\
\left\{ v_{i},v_{i+1}\right\} &=&v_{i}v_{i+1}\left( v_{i}+v_{i+1}\right) 
\ , i=2,\ldots ,n-3  \nonumber \\
\left\{ v_{n-2},v_{n-1}\right\} &=&v_{n-2}v_{n-1}\left(
2v_{n-1}+v_{n-2}\right)  \nonumber \\
\left\{ v_{n-1},v_{n}\right\} &=&2v_{n-1}v_{n}\left( v_{n}-v_{n-1}\right) 
\nonumber \\
\left\{ v_{i},v_{i+2}\right\} &=&v_{i}v_{i+1}v_{i+2}\ , i=1,\ldots ,n-3
\nonumber \\
\left\{ v_{n-2},v_{n}\right\} &=&v_{n-2}v_{n}\left( v_{n-2}+2v_{n}\right) 
\nonumber \\
\left\{ v_{n-3},v_{n}\right\} &=&v_{n-3}v_{n-2}v_{n}\ ;  \nonumber
\end{eqnarray}%

all other brackets are zero. As in the case of KM system we suppose that $n$
is odd $\left( n=2m+1\right) $ and we look  for a bracket $\pi _{1}$
which satisfies $\pi _{3}\nabla H_{2}=\pi _{1}\nabla H_{4}.$

We define 
\bd
\tau _{ij}=-\tau _{ji}=v_{2i-1}\prod_{k=i}^{j-1}\frac{v_{2k+1}}{v_{2k}}\ \ 
{\rm for}\ i<j\ \ ,\ \ \tau _{ii}=v_{2i-1}\ ,
\ed
and we define the bracket  $\pi _{1}$  as follows:
\begin{eqnarray}
\left\{ v_{i},v_{j}\right\} &=&\left( -1\right) ^{i+j-1}\tau _{\left[ \frac{i%
}{2}\right] +1,\left[ \frac{j+1}{2}\right] }\ \ {\rm  for \ }1\leq i<j\leq
n-2\ , \label{a44} \\
\left\{ v_{i},v_{n-1}\right\} &=&\left\{ v_{i},v_{n}\right\} =\frac{\left(
-1\right) ^{i+n}}{2}\tau _{\left[ \frac{i}{2}\right] +1,\left[ \frac{n}{2}%
\right] }\ \ {\rm for  }\ i=1,\ldots ,n-2\ ,  \nonumber \\
\left\{ v_{n-1},v_{n}\right\} &=&-\left\{ v_{n},v_{n-1}\right\} =\frac{1}{2}%
\left( v_{n}-v_{n-1}\right) \ .  \nonumber
\end{eqnarray}

To illustrate, we give the Poisson matrix of the bracket $\pi _{1}$ in the
case $n=7.$

\[
\pi _{1}=\left[ 
\begin{array}{ccccccc}
0 & \tau _{11} & -\tau _{12} & \tau _{12} & -\tau _{13} & \frac{1}{2}\tau
_{13} & \frac{1}{2}\tau _{13} \\ 
&  &  &  &  &  &  \\ 
-\tau _{11} & 0 & \tau _{22} & -\tau _{22} & \tau _{23} & -\frac{1}{2}\tau
_{23} & -\frac{1}{2}\tau _{23} \\ 
&  &  &  &  &  &  \\ 
\tau _{12} & -\tau _{22} & 0 & \tau _{22} & -\tau _{23} & \frac{1}{2}\tau
_{23} & \frac{1}{2}\tau _{23} \\ 
&  &  &  &  &  &  \\ 
-\tau _{12} & \tau _{22} & -\tau _{22} & 0 & \tau _{33} & -\frac{1}{2}\tau
_{33} & -\frac{1}{2}\tau _{33} \\ 
&  &  &  &  &  &  \\ 
\tau _{13} & -\tau _{23} & \tau _{23} & -\tau _{33} & 0 & \frac{1}{2}\tau
_{33} & \frac{1}{2}\tau _{33} \\ 
&  &  &  &  &  &  \\ 
-\frac{1}{2}\tau _{13} & \frac{1}{2}\tau _{23} & -\frac{1}{2}\tau _{23} & 
\frac{1}{2}\tau _{33} & -\frac{1}{2}\tau _{33} & 0 & \frac{1}{2}\left(
v_{7}-v_{6}\right) \\ 
&  &  &  &  &  &  \\ 
-\frac{1}{2}\tau _{13} & \frac{1}{2}\tau _{23} & -\frac{1}{2}\tau _{23} & 
\frac{1}{2}\tau _{33} & -\frac{1}{2}\tau _{33} & -\frac{1}{2}\left(
v_{7}-v_{6}\right) & 0%
\end{array}%
\right] 
\]%
\begin{eqnarray*}
&& \\
&=&\left[ 
\begin{array}{ccccccc}
0 & v_{1} & -\frac{v_{1}v_{3}}{v_{2}} & \frac{v_{1}v_{3}}{v_{2}} & -\frac{%
v_{1}v_{3}v_{5}}{v_{2}v_{4}} & \frac{v_{1}v_{3}v_{5}}{2v_{2}v_{4}} & \frac{%
v_{1}v_{3}v_{5}}{2v_{2}v_{4}} \\ 
&  &  &  &  &  &  \\ 
-v_{1} & 0 & v_{3} & -v_{3} & \frac{v_{3}v_{5}}{v_{4}} & -\frac{v_{3}v_{5}}{%
2v_{4}} & -\frac{v_{3}v_{5}}{2v_{4}} \\ 
&  &  &  &  &  &  \\ 
\frac{v_{1}v_{3}}{v_{2}} & -v_{3} & 0 & v_{3} & -\frac{v_{3}v_{5}}{v_{4}} & 
\frac{v_{3}v_{5}}{2v_{4}} & \frac{v_{3}v_{5}}{2v_{4}} \\ 
&  &  &  &  &  &  \\ 
-\frac{v_{1}v_{3}}{v_{2}} & v_{3} & -v_{3} & 0 & v_{5} & -\frac{v_{5}}{2} & -%
\frac{v_{5}}{2} \\ 
&  &  &  &  &  &  \\ 
\frac{v_{1}v_{3}v_{5}}{v_{2}v_{4}} & -\frac{v_{3}v_{5}}{v_{4}} & \frac{%
v_{3}v_{5}}{v_{4}} & -v_{5} & 0 & \frac{v_{5}}{2} & \frac{v_{5}}{2} \\ 
&  &  &  &  &  &  \\ 
-\frac{v_{1}v_{3}v_{5}}{2v_{2}v_{4}} & \frac{v_{3}v_{5}}{2v_{4}} & -\frac{%
v_{3}v_{5}}{2v_{4}} & \frac{v_{5}}{2} & -\frac{v_{5}}{2} & 0 & \frac{%
v_{7}-v_{6}}{2} \\ 
&  &  &  &  &  &  \\ 
-\frac{v_{1}v_{3}v_{5}}{2v_{2}v_{4}} & \frac{v_{3}v_{5}}{2v_{4}} & -\frac{%
v_{3}v_{5}}{2v_{4}} & \frac{v_{5}}{2} & -\frac{v_{5}}{2} & -\frac{v_{7}-v_{6}%
}{2} & 0%
\end{array}%
\right]\ .
\end{eqnarray*}

\bigskip

The following theorem is from \cite{kouzaris}.

\begin{theorem}
$\left( i\right) $ $\pi _{1}$ , $\pi _{3}$ are Poisson.\newline
$\left( ii\right) $ The function 
\[
\frac{1}{4}H_{2}=\frac{1}{8}Tr(L^{4})=v_{n-2}v_{n}+2v_{n-1}v_{n}+%
\sum_{i=1}^{n-2}v_{i}v_{i+1}+\frac{1}{2}\sum_{i=2}^{n-2}v_{i}^{2} 
\]%
is the Hamiltonian of \ the $BV$ $D_{n+1}$ system\ with respect to the
bracket $\pi _{1}.$\newline
$\left( iii\right) $ The function 
\[
F=\left( v_{n}-v_{n-1}\right) \prod_{i=1}^{n-2}v_{i}\  ,
\]%
is the Casimir of \ the $BV$ $D_{n+1}$ system\ in the  bracket $\pi _{1}.$%
\newline
$\left( iv\right) $ $\pi _{1}$ , $\pi _{3}$ are compatible.\newline

$\left( v\right) $  $\pi _{3}\nabla H_{2}=\pi _{1}\nabla H_{4}$ \ .
\end{theorem}

\section{From Volterra to Birkhoff}

We consider the Volterra $D_{n+1}$ system (\ref{a40}).
We assume that $n$ is odd,  equal to  $2m +1$ and rename again the variables (i.e. use $u_k$ in place of
$v_k$). We recall the equations for the system:

\begin{eqnarray}
\dot{u}_{1} &=&u_{1}\left( u_{1}+ u_{2}\right)  \label{a11} \\
\dot{u}_{k} &=&u_{k}(u_{k+1}-u_{k-1}) \ \  2\leq k\leq n-3  \nonumber \\
\dot{u}_{n-2}&=&u_{n-2}\left( u_{n}+u_{n-1}-u_{n-3}\right) ,  \nonumber \\
\dot{u}_{n-1} &=&u_{n-1}\left( u_{n}-u_{n-1}-u_{n-2}\right) , \nonumber \\
\dot{u}_{n}&=&-u_{n}\left( u_{n}-u_{n-1}+u_{n-2}\right) \ . \nonumber
\end{eqnarray}

We make the transformation (the analogue of H\'enon transformation (\ref{a25}) for the KM-system)

\begin{eqnarray}
a_{1} &=&\frac{i}{2}\left( u_{n}-u_{n-1}\right) \label{a12} \\
a_{j} &=&\frac{1}{2}{\displaystyle \sqrt{u_{n-2j+2}u_{n-2j+1}} } \ \ \ \  j=2,3,\ldots ,m \nonumber \\
a_{m+1} &=&\frac{i}{2}u_{1} \nonumber \\
b_{1} &=&-\frac{1}{2}(u_{n}+u_{n-1}+u_{n-2}) \nonumber \\
b_{j} &=&-\frac{1}{2}(u_{n-2j+1}+u_{n-2j}) \ \ \ \ j=2,3,\ldots ,m. \nonumber
\end{eqnarray}

This transformation is derived  by mimicking the  construction of Moser which takes the $A_n$ Volterra lattice  to the
Toda lattice, i.e. the  construction  that was described in the introduction. The formulas  (\ref{a12}) are obtained by considering 
$L^2$, where $L$ is given by (\ref{a41}), and assigning suitable entries to the variables $a_i$, $b_i$.

We calculate:

\begin{eqnarray*}
\dot{a}_{1} &=&\frac{i}{2}\left( \dot{u}_{n}-\dot{u}_{n-1}\right) \\
&=&\frac{i}{2}\left[
-u_{n}(u_{n}-u_{n-1}+u_{n-2})-u_{n-1}(u_{n}-u_{n-1}-u_{n-2})\right] \\
&=&\frac{i}{2}(-u_{n}^{2}-u_{n-2}u_{n}+u_{n-1}^{2}+u_{n-2}u_{n-1}) \\
&=&\frac{i}{2}(u_{n-1}-u_{n})(u_{n}+u_{n-1}+u_{n-2}) \\
&=&2a_{1}b_{1}.
\end{eqnarray*}

In a similar fashion we obtain the equations of  motion in the new variables $(a_i, b_i)$.

\begin{eqnarray}
\dot{a}_{1} &=&2a_{1}b_{1} \label{a13}  \\
\dot{a}_{i} &=&a_{i}\left( b_{i}-b_{i-1}\right)  \ \ i=2,3,\ldots ,m \nonumber \\
\dot{a}_{m+1} &=&-2a_{m+1}b_{m} \nonumber \\
\dot{b}_{i} &=&2\left( a_{i+1}^{2}-a_{i}^{2}\right)  \    i=1,2,\ldots ,m \  .   \nonumber
\end{eqnarray}

This system can be written in a Lax pair form as follows:

\be
L =\pmatrix{b_{1} & a_{1} & a_{2} & 0 & \cdots  & \cdots  & 0 \cr
a_{1} & -b_{1} & 0 & -a_{2} & \ddots  &  & \vdots  \cr
a_{2} & 0 & b_{2} & 0 & \ddots  & \ddots  & \vdots   \cr
0 & -a_{2} & 0 & -b_{2} & \ddots  & a_{m} & 0  \cr
\vdots  & \ddots  & \ddots  & \ddots  & \ddots  & 0 & -a_{m}  \cr
\vdots  &  & \ddots  & a_{m} & 0 & b_{m} & a_{m+1}  \cr
0 & \cdots  & \cdots  & 0 & -a_{m} & a_{m+1} & -b_{m} }
 \ , \label{a14} 
\ee

\bd
B=\pmatrix{0 & -a_{1} & a_{2} & 0 & \cdots  & \cdots  & 0 \cr
a_{1} & 0 & 0 & a_{2} & \ddots  &  & \vdots  \cr
-a_{2} & 0 & 0 & \ddots  & \ddots  & \ddots  & \vdots  \cr
0 & -a_{2} & \ddots  & \ddots  & \ddots  & a_{m} & 0 \cr
\vdots  & \ddots  & \ddots  & \ddots  & 0 & 0 & a_{m} \cr
\vdots  &  & \ddots  & -a_{m} & 0 & 0 & a_{m+1} \cr
0 & \cdots  & \cdots  & 0 & -a_{m} & -a_{m+1} & 0} \ .
\ed

We have that $\dot{L}=[B, L]$  is equivalent to equations (\ref{a13}).

The functions 
\bd
H_{2k}={ 1 \over 2k} {\rm tr} \,   L^{2k} \ , \ \ k=1,2, \dots, m \ .
\ed
are independent constants of motion.
 There is also 
a Casimir, which makes the system completely integrable, but these  functions are enough for our purpose.

We take  $H_2$ as  the  Hamiltonian and define a Poisson bracket $\pi_1$ as follows:
We consider the mapping from ${\bf R}^{2m+1}$ to ${\bf R}^{2m+1}$ given by  (\ref{a12}). In other 
words it is the mapping which transforms the $u_i$ to the new variables $(a_i, b_i)$. The image of the
bracket (\ref{a44})   (up to a constant multiple) is given by

\begin{eqnarray}
\left\{ a_{1},b_{1}\right\} &=&a_{1} \label{a15} \\
\left\{ a_{i},b_{i}\right\} &=&\frac{1}{2}a_{i} \ \ \ i=2,3,\ldots ,m \nonumber\\
\left\{ a_{i+1},b_{i}\right\} &=&-\frac{1}{2}a_{i} \ \ \ \ i=1,2,\ldots m-1 \nonumber \\
\left\{ a_{m+1},b_{m}\right\} &=&-a_{m+1};  \nonumber
\end{eqnarray}

all other brackets are zero.  The function 
\bd
C=a_1 a_2^2 a_3^2 \dots a_m^2 a_{m+1} 
\ed
 is a Casimir.  This is the Casimir $F$  of theorem 2 in $(a,b)$ coordinates.

We also have involution of invariants,  $ \{  H_i, H_j \}=0$. 
  We   denote this bracket by $\pi_1$.  
We have
\bd
\pi_1 \nabla H_2 
\ed
is equivalent to equations (\ref{a13}).

Define a Hamiltonian system in ${\bf R}^{2m}$ with coordinates $(q_1, \dots, q_m, \,  p_1, \dots, p_m)$
 by

\be
H(q,p)=\frac{1}{2}\sum_{j=1}^{m}p_{j}^{2}+
\sum_{j=1}^{m-1}e^{q_{j}-q_{j+1}} +e^{-2q_{1}}+e^{2q_{m}}. \label{a16}
\ee

As in the case of the classical Toda lattice we make a Flaschka--type transformation

\begin{eqnarray}
a_{1} &=&\frac{1}{\sqrt{2}}e^{-q_{1}}, \ \ \ a_{m+1}=\frac{1}{\sqrt{2}}
e^{q_{m}} \label{a17} \\
a_{i} &=&\frac{1}{2}e^{\frac{1}{2}\left( q_{i-1}-q_{i}\right) } \ \ i=2,3,\ldots ,m \nonumber \\
b_{i} &=&-\frac{1}{2}p_{i} \ \ i=1,2,\ldots ,m \ .  \nonumber
\end{eqnarray}

This is a mapping from ${\bf R}^{2m} \to {\bf R}^{2m +1} $. Note that we are not using  transformation (\ref{a66}) but a more traditional variation.

We easily verify that Hamilton's equations for the variables $(a_j, b_j)$ are precisely equations
(\ref{a13}).  For example,
\bd
\dot{a}_1 =-{1\over \sqrt{2}} e^{-q_1} \dot{q}_1=-a_1  { \p H \over \p p_1} =-a_1 p_1=-a_1 (-2 b_1)=2 a_1 b_1
\ .
\ed
We recall that the system (\ref{a13}) has  $H_2, H_4, \dots, H_{2m}$  as a set of integrals in involution.
 Reverting back to 
the original variables $(q_i,\, p_i)$ in ${\bf R}^{2m}$ we obtain $m$  independent integrals in involution,
and this proves the integrability of (\ref{a16}).  For example 
\bd
H_2=\sum_{i=1}^m b_i^2 +
a_1^2+2 \sum_{i=2}^m a_i^2  +a_{m+1}^2 
\ed
corresponds to the Hamiltonian (\ref{a16}). The integrals are of degrees $2,4,6, \dots, 2m$ in the momenta. The
Casimir $C$ reduces to a constant equal to ${1 \over 2^{2m-1}}$.

We  note that for  the system (\ref{a13}) we may   define a cubic bracket $\pi_3$ which
satisfies the Lenard relation 

\begin{equation}
 \pi_3 \ \nabla \ H_2 = \pi_1 \ \nabla  \ H_4  \ .  \label{a18}
\end{equation}

This bracket is the image of the bracket (\ref{a43}) under  the mapping (\ref{a12}).
The bracket $\pi _{3}$ is given by
\begin{eqnarray*}
\left\{ a_{i},a_{i+1}\right\}  &=&a_{i}a_{i+1}b_{i} \ \ \ 
i=2,3,\ldots ,m-1 \\
\left\{ a_{i},a_{i+1}\right\}  &=&2a_{i}a_{i+1}b_{i}  \ \ \  i=1 \ \ \
{\rm and }\ \ m \\
\left\{ b_{i},b_{i+1}\right\}  &=&2a_{i+1}^{2}(b_{i}+b_{i+1}) \ \ \
i=1,2,\ldots ,m-1 \\
\left\{ a_{1},b_{1}\right\}  &=&2a_{1}(a_{1}^{2}+b_{1}^{2}) \\
\left\{ a_{i},b_{i}\right\}  &=&a_{i}(a_{i}^{2}+b_{i}^{2}) \ \ 
i=2,3,\ldots ,m-1 \\
\left\{ a_{m},b_{m}\right\}  &=&a_{m}(a_{m}^{2}+b_{m}^{2}-a_{m+1}^{2}) \\
\left\{ a_{1},b_{2}\right\}  &=&2a_{2}^{2}a_{1} \\
\left\{ a_{i},b_{i+1}\right\}  &=&a_{i+1}^{2}a_{i}  \ \ \  i=2,3,\ldots
,m-1 \\
\left\{ a_{2},b_{1}\right\}  &=&-a_{2}(a_{2}^{2}+b_{1}^{2}-a_{1}^{2}) \\
\left\{ a_{i+1},b_{i}\right\}  &=&-a_{i+1}(a_{i+1}^{2}+b_{i}^{2}) \ \
i=2,3,\ldots ,m-1 \\
\left\{ a_{m+1},b_{m}\right\}  &=&-2a_{m+1}(a_{m+1}^{2}+b_{m}^{2}) \\
\left\{ a_{i+2},b_{i}\right\}  &=&-a_{i+1}^{2}a_{i+2} \ \ \ 
i=1,2,\ldots ,m-2 \\
\left\{ a_{m+1},b_{m-1}\right\}  &=&-2a_{m}^{2}a_{m+1} \ ;
\end{eqnarray*}
all other brackets are zero. It is clear, by the way it was constructed, that this  bracket is Poisson,
  compatible with $\pi_1$ and that  the integrals $H_{i}$ are all  in involution.
We close with a few  remarks:

\noindent
\begin{enumerate}
\item
One could try to obtain  a bi--Hamiltonian formulation of the system 
(\ref{a13}) following the recipe  of \cite{damianou2}.
The basic steps in the construction are the following:
Write the second bracket $\pi_3$  in $(q,p)$ coordinates (call it $J_3$) and define a recursion operator
in $(q,p)$ space by   inverting the standard symplectic bracket ($J_1$).
 Define the negative recursion operator in $(q,p)$ space by inverting the second bracket $J_3$.
 Define a new rational bracket
$J_{-1}$ by $J_{-1}=J_1 J_3^{-1}J_1$. 
Finally,  project the bracket $J_{-1}$ into the $(a,b)$ space to obtain a rational bracket $\pi_{-1}$.  The result is
a bi--Hamiltonian formulation of the system:
\bd
\pi_1 \nabla H_2 = \pi_{-1} \nabla H_4  \ .
\ed
Note that a recursion operator in the $(a,b)$ space  does not exist since $\pi_1$ and $\pi_3$ are both 
singular. So far, we were unable to compute the bracket $J_3$ and we are not certain that it can be computed.

\item
It is a straightforward recursive process to solve the $u_i$ as functions of the $a_i$, $b_i$ using
(\ref{a12}). After substitution of the values of $a_i$, $b_i$ from (\ref{a17}) we obtain an expression of the
$u_i$ as functions of $q_i$, $p_i$. The formulas are too complicated but in principle the invariants, symmetries and
higher Poisson brackets  in the $(q,p)$ space of (\ref{a16}) transfer 
to the corresponding ones for the  Volterra system (\ref{a12}) via this mapping.

\item
One may predict the degrees of the integrals by computing the  Kovalevskaya exponents as in  \cite{emelyanov2}. In our
case the degrees of the invariants $2,4, \dots $ agree with the predicted values.

\item
The Casimir
\bd
C=a_1 a_2^2 a_3^2 \dots a_m^2 a_{m+1} 
\ed
may be obtained in a different way using the following observation:
 The Casimir is the product of all the $a_i$ raised to certain exponents.
 We note that the exponents (not to be confused with the Kovalevskaya exponents)
$(1,2,2, \dots, 2, 1)$   can be determined from the condition
\bd
{\bf v}_1+2 {\bf v}_2+ \dots+ 2 {\bf v}_{N-1}+ {\bf v}_N =0 \ .
\ed
This is always the case for systems of the form (\ref{a1}) if  $N >n$.

\item
The procedure of this paper works equally well for the system
\be
  H(q_1, \dots, q_n, \,  p_1, \dots, p_n) = \sum_{i=1}^n \,  { 1 \over 2} \, p_i^2 + 
\sum _{i=1}^{n-1} \,  e^{ q_i-q_{i+1}}  + \alpha e^{-2 q_1}+  + \beta e^{q_n}
 \ .
\ee
The only difference is that we use the  Volterra $D_{n+1}$ system  with $n=2m$.

\item
The Hamiltonian (\ref{a16}) is positive, and the  Casimir $C$ keeps the variables $a_k$ from becoming zero. Thus, the energy surface is compact, and solutions lie on
tori. This suggests that it should be possible to introduce a spectral parameter into the Lax pair and so get to Riemann surfaces and theta functions.
\end{enumerate}

\smallskip
\noindent
{\bf Acknowledgments} We thank the anonymous referee  for  useful remarks and corrections which 
made the paper more readable and precise.

\end{document}